# Magnetic contrast layers with functional SiO$_2$ coatings for soft matter studies with polarised neutron reflectometry

O. Dikaia[1*], A. Luchini[2], T. Nylander[3,4], A. Grunin[1], A. Vorobiev[5] and A. Goikhman[1]

[1] Königssystems UG, 22869 Schenefeld, Germany

[2] Department of Physics and Geology, University of Perugia, via Alessandro Pascoli, 06123 Perugia, Italy

[3] NanoLund and Physical Chemistry, Lund University, 22100 Lund, Sweden

[4] LINXS Institute of Advanced Neutron and X-ray science, Scheelevägen 19, Lund, 22370, Sweden

[5] Department for Physics and Astronomy, Uppsala University, 75120 Uppsala, Sweden

*Correspondence to: olga.dikaia@koenigssystems.de, olga.a.dikaya@gmail.com

## Abstract

This study introduces silicon substrates with a switchable magnetic contrast layer (MCL) for polarised neutron reflectometry experiments (PNR) at solid/liquid interface to study soft matter surface layers. The advantage with neutron reflectometry (NR) data is that structural and compositional information can be enhanced by using different isotopic contrast on the same sample. This approach is normally referred to as contrast matching, which can be achieved by using solvents with different isotopic contrast, e. g. different H$_2$O/D$_2$O ratio, and/or by selective deuteration of the molecules. However, some soft matter system might be perturbed by this approach, or it might not be possible, particularly for biological samples. In this scenario, solid substrates with a MCL are an appealing alternative, as the magnetic contrast with the substrate can be used to partially recover the information on the sample structure. More specifically, in this study, a magnetically soft Fe layer coated with SiO$_2$ was produced by ion-beam sputter deposition on silicon substrates of different sizes. The structure was evaluated using X-ray reflectometry (XRR), atomic force microscopy (AFM), vibrating sample magnetometry (VSM), and PNR. The collected data showed high quality and repeatability of the MCL parameters, regardless of the substrate size and thickness of the capping SiO$_2$ layers. As compared to other kinds of substrates with MCL layer previously proposed, which used Au capping layer, the SiO$_2$ capping layer allows to reproduce the typical surface of standard silicon substrate used for NR experiment and compatible with a large variety of soft matter samples. We demonstrated such application, by using ready-to-measure 50×50×10 mm$^3$ substrates in PNR experiments for the characterisation of a lipid bilayer in a single solvent contrast. Overall, the article highlights the potential of PNR with MCL for investigation of soft matter samples.

**Introduction**

Neutron reflectometry (NR) technique is a powerful tool for non-invasively investigating complexly ordered soft matter systems at solid/liquid interfaces. However, the solutions of the inverse scattering equations are not unique. To increase reliability of the results, multiple NR measurements in combination with characterization by axillary experimental methods are desired. A commonly utilised approach is to collect reflectivity curves from structurally identical but isotopically distinct samples. Deuteration is the most applicable technique for this purpose, wherein protium is fully or partially replaced with deuterium [1–3]. In particular, solvent contrast exchange is a common protocol during NR experiments at solid/liquid interface, which consists in collecting data on the same sample after a sequential replacement of the aqueous solvent with mixtures of $H_2O$ and $D_2O$ at variable ratios [4]. These results in several data sets referring to the same sample but in contact with isotopically distinct solvents which corresponds to different contrasts, i. e. difference between the sample's neutron scattering-length density (SLD) and the solvent SLD. The simultaneous analysis of the data collected in different contrasts, typically at least three, provides the necessary information to determine a unique solution to the sample structure.

Although solvent contrast exchange is a well-established approach, there are some specific cases where it is not applicable. Some soft matter samples can be affected by the solvent flow within the NR cell needed for the solvent replacement, either in terms of structural changes or by being (partially) removed from the substrate. As an example, a previous NR study on the interaction between lipid bilayers and a soluble protein, e. g. a-synuclein, required the injection of new protein solution during the solvent exchange as the initially absorbed protein molecules would be otherwise removed during the solvent rinsing [5]. Although, this approach produced reasonable information on the sample structure, it is quite evident that the assumption about the sample structure being identical in the different contrasts would not hold for this case. A similar issue was found also in another study, where in the end the author renounced to use NR as characterisation techniques [6]. There are several other cases, currently not accessible with NR because of the instability of the sample during solvent replacement.

An alternative approach is to investigate soft matter samples by employing a solid substrate with a magnetic contrast layer (MCL). In such experiment polarized neutron beam interacts with MCL magnetized by permanent magnetic field. Depending on the orientation of the neutron polarization – parallel or antiparallel to the MCL magnetization – two different reflectivity curves ($R^+$ and $R^-$ respectively) can be obtained [7]. Both resulting SLD profiles correspond to the same soft matter sample and differ by a single parameter – magnetic SLD of the MCL. Magnetically soft layers have been already successfully used for the investigation of soft matter samples, such as a wide range of biological membranes and polymer layers [8–12]. However, in these experiments a magnetic layer after deposition on a silicon substrate was either capped with a gold layer producing the outermost surface for the soft matter sample deposition [8–10] or the magnetic layer remained uncovered at all, without functionalised coating applied [11]. But in fact, a silicon oxide capping layer is a preferable solution for the investigation of various soft matter systems, such as e. g. lipid bilayers. [13]. To the best of our knowledge, only one previously cited PNR study utilised a $SiO_2$ top layer to protect the MCL [12].

Nevertheless, growing a smooth (low roughness) top oxide layer while avoiding degradation of the magnetic layer underneath remained so far, a significant challenge. This work describes a set of ready-to-use substrates for PNR experiments on soft matter samples that consist of a thin iron film as the MCL deposited on a silicon crystal and capped with a $SiO_2$ layer of different thicknesses and a few Å roughness. Ion-beam sputter deposition (IBSD) was used to grow 10 nm Fe layers in the same vacuum process as the SiO$_2$ capping layer. The structure and composition of the produced substrates were characterized by atomic force microscopy (AFM), X-ray reflectometry (XRR) and vibrating sample magnetometry (VSM). In addition, PNR data was collect at the Super ADAM reflectometer [14]. We demonstrated that magnetic properties of the MCL are independent of the design of the top SiO$_2$ layer. Finally, we chose one of the produced substrates – the one with a 10 nm thick SiO$_2$ layer – for testing the deposition and characterisation of a lipid bilayer. During preliminary tests we found out that this thickness of the SiO$_2$ layer provides the best compatibility with saline solution that are required for the investigation of lipid bilayers. In this study, we investigated the structure of a 1-palmitoyl-2-oleoyl-glycerol-3-phosphocholine(POPC)/1-palmitoyl-2-oleoyl-sn-glycero-3-phospho-(1'-rac-glycerol)(POPG)/ergosterol(Erg) bilayer in physiological buffer, which represents a relevant model system of the yeast plasma membrane and provides a suitable platform to test the impact of antifungal drugs. Altogether, we show that our substrates with MCL and SiO$_2$ capping layer are suitable for the formation and characterisation of such supported lipid bilayer in a single solvent contrast.

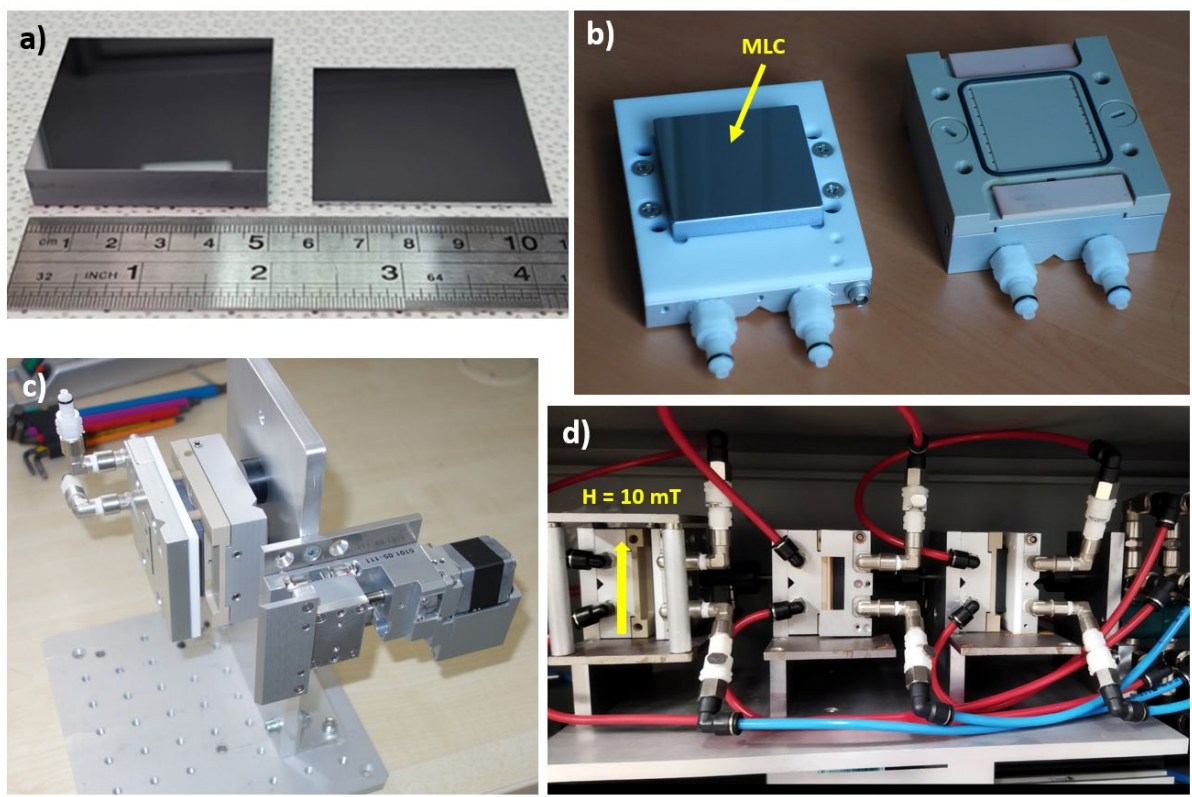

Figure 1. a) Typical substrates with dimensions of 50×50×10 and 50×50×1 mm$^3$, b) disassembled experimental cell for solid-liquid interface study, c) assembled cell mounted on a support with motorized beam stop, d) motorized sample changer for three cells in the external magnetic field system (only left cell).

**Experimental**

*Substrates production*

All substrates were produced by sequential deposition of individual Fe and SiO$_2$ thin layers by the IBSD. The deposition system was especially designed for optical coatings that demand a high thickness uniformity grade over large areas [15]. As a result, it allows growing of thin films with a suitable surface area for neutron reflectometry experiments. A collimated Ar-beam was used to sputter Fe and SiO$_2$ targets of high purity to form smooth layers with a fine interface. The 12 cm Kaufman ion source was set to a beam energy 1000 eV for Fe and 1000/800 eV for SiO$_2$. No additional heating or post-processing was applied for these basic structural components. As substrates we used 50×50 mm$^2$ silicon blocks of thickness 10 mm (additional test samples deposited on 0.4 mm, 1 mm and 3 mm Si wafers are presented in SI). MCLs deposited on 10 mm substrates can be directly used in a special experimental cell (Fig. 1) designed for NR measurements at solid-liquid interface. For better recognition, the numbers in substrate names correspond to the SiO$_2$ top layer thickness in nm. In this paper we focus on the following structures listed in Table 1.

**Table 1.** List of substrates.

| Substrate ID | Fe thickness, nm | SiO$_2$ thickness, nm | Si block thickness, mm | area, mm$^2$ |
|---|---|---|---|---|
| S-*3.5* | 10 | *3.5* | 10 | 2500 |
| S-*10* | 10 | *10* | 10 | 2500 |
| S-*80* | 10 | *80* | 1 | 2500 |

*Substrate characterisation*

Surface roughness was initially assessed on test samples with smaller sizes compared to those used for PNR measurements. We collected AFM data with at the Aist SmartSPM-1000 setup with an NSG10 (125 μm length) cantilever with curvature radius of 10 nm in a tapping mode. Test samples grown in the same process were requires as the largest substrates were not compatible with our laboratory equipment. We collected maps from 10×10 μm$^2$ and 2×2 μm$^2$ areas for each sample. All AFM images were processed with Gwyddion software [16]. It is worth noticing that the Gwyddion software calculates the Rms parameter by averaging local values. Thus, the *ab initio* topography of substrates may be neglected in the evaluation of parameters of thin films. XRR curves were obtained by the Bruker D8 Discover diffractometer (Cu-Kα, λ=0.154 nm) and refined via the commercial software Diffrac.Leptos. A preliminary VSM investigation was conducted at the LakeShore 7404 magnetometer at five angular positions in the film plane (0º-180º) for S-10 and at one position for other samples.

*Polarised neutron reflectometry*

PNR experiments were carried out on neutron reflectometer Super ADAM [14] (Institut Laue-Langevin, Grenoble, France) using a monochromatic neutron beam with wavelength λ=5.2 Å and polarisation P$_0$=99.8%. A detailed description of the polarised reflectometry technique can be found elsewhere [17]. Reflectivity data sets were collected at room temperature in the external magnetic field of B=10 mT created by a system of permanent magnets.

The experimental data was reduced using pySAred software (provided by ILL) and fitted using the GenX package [18]. PNR datasets R$^+$ and R$^-$ (corresponding to the neutron beam polarization parallel and antiparallel to the magnetisation of Fe layer) were fitted simultaneously, providing nuclear and magnetic components of SLD (SLD$_n$ and SLD$_m$ respectively) as a function of a distance z from the surface of the Si substrate.

In the experiment with a lipid bilayer, the S-10 structure was used. PNR measurements were performed for the sample either in D$_2$O or in H$_2$O. Sample was sealed in a solid-liquid cell (Fig. 2) connected to a high-performance liquid chromatography pump, simultaneously allowing solvent exchange and sample injection.

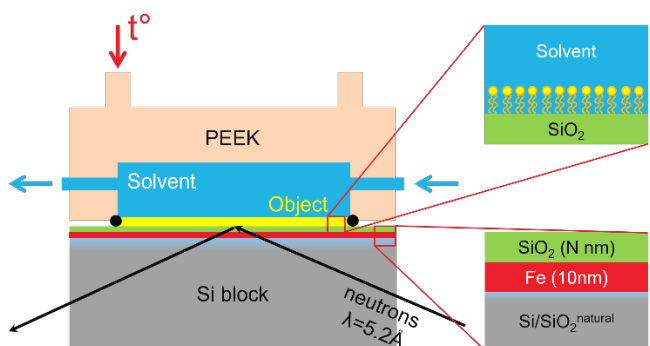

Figure 2. The sketch of the sample cell for solid-liquid interface study with the experimental details. Schematic structures of magnetic-capping layers on the substrate and model structure are presented in insets.

*Supported lipid bilayer*

A supported lipid bilayer (SLB) was prepared by vesicle fusion [4]. Lipid vesicles with composition POPC/POPG/Erg (60/30/10 mol/mol/mol) were prepared by dissolving a lipid film containing the lipids at the above-mentioned ratio with a 0.15 M $KNO_3$ solution in $D_2O$. Preliminary test measurements highlighted that chlorine ions in the buffer could compromise the MCL layer. On the other hand, 0.15 M $KNO_3$ did not affect the MCL support surface and improved the formation of the SLB. The lipid suspension was sonicated for 5 min with a tip sonicator before injection in the NR solid/liquid cell. After approximately 30 min, the cell was rinsed with 20 ml of 0.15 M $KNO_3$ $D_2O$ solution in order to remove the vesicle excess and favour vesicle fusion on the substrate surface, which led to the formation of the SLB. PNR curves were collected using the same settings as for the measurements on the bare substrates. The data were analysed using the Motofit software [19]. The parameters for the substrate structure were kept fixed to those determined during the characterisation of the bare substrate, and only those referring to the lipid bilayer were optimised to the experimental data. Data corresponding to $R^+$ and $R^-$ were analysed simultaneously with the same sample structure.

**Results and discussion**

*Analysis of grown substrates*

Values of structural parameters for all three substrates used as obtained by AFM and XRR are collected in Table 2. Although roughness of the topmost surface for all substrate is very low, but the S-80 surface looks exceptionally smooth with Rms=0.2 nm (see in Fig. S1). We assume, this smoothing effect has caused by amorphous nature of the films grown by ion-beam sputtering deposition at the room temperature and was reported by other groups for reactive [20,21] and non-reactive [22] $SiO_2$ deposition. In both cases, thick (>40 nm) $SiO_2$ layers demonstrated roughness less than with 0.2 nm. For thin layers (<40 nm) the situation is not so obvious. In the IBSD method, initial stage of thin film growth goes according to Volmer-Weber mechanism [23]. Following this theory, growth islands after coalescence form films with well-defined granular structure. An exact film thickness, where coalescence of growth sites happens, depends on growth parameters, but this granular structure easily can be seen in $SiO_2$ films with low thickness (3.5 and 10 nm) by AFM. Also, films of thickness comparable with substrate roughness have morphology repeating irregularities of substrate surface, it may be defined as "surface morphology inheritance". In Fig. S1.1 and S1.2 which were obtained from test samples (for S-3.5 and S-10 correspondingly) both effects are presented. A higher roughness of 3.5 and 10 nm $SiO_2$ can be explained as synergy of granular structure and "morphology inheritance" from Si blocks.

**Table 2.** Structural parameters of MCL substrates as obtained from AFM and XRR data.

| Substrate ID | AFM | | XRR | | | | | |
| --- | --- | --- | --- | --- | --- | --- | --- | --- |
| | $Rms_{10\times10\mu m}$, nm | $Rms_{2\times2\mu m}$, nm | Fe | | | SiO$_2$ | | |
| | | | $d_{Fe}$, nm | $\sigma_{Fe}$, nm | $\rho_{Fe}$, g/cm$^3$ | $d_{SiO2}$, nm | $\sigma_{SiO2}$, nm | $\rho_{SiO2}$, g/cm$^3$ |
| S-3.5 | 0.3 | 0.2 | 10.6 | 0.4 | 7.9 | 3.7 | 0.5 | 2.5 |
| S-10 | 0.3 | 0.3 | 10.5 | 0.4 | 8.0 | 11.1 | 0.7 | 2.3 |
| S-80 | 0.2 | 0.2 | 9.6 | 0.2 | 7.9 | 79.9 | 0.6 | 2.6 |

An analysis based on X-ray reflectometry approves compliance with the thicknesses of layers applied to the quartz crystal microbalance during deposition. As the IBSD films grown at room temperature are amorphous, a small amount of dissolved Ar can slightly change mass density ($\rho$) of Fe and SiO$_2$ layers, so it may differ from bulk values. Therefore, in the fitting procedure we allowed these parameters to be varied. The experimental XRR data curves are shown in Fig. 3 together with best-fit model curves and corresponding density profiles. We have found that Fe layer of all samples has thickness $d_{Fe}$=10.2±0.5 nm and density $\rho_{Fe}$=7.9±0.1 g/cm$^3$, which is a in very good agreement with the value of the bulk iron (7.874 g/cm$^3$). The thicknesses of SiO$_2$ layers vary due to deposition models and have a deviation of less than 5.7% of the target value. The density of SiO$_2$ layers $\rho_{SiO2}$=2.5±0.1 g/cm$^3$ slightly lower than nominal bulk value of 2.65 g/cm$^3$. This implies that amorphous SiO$_2$ contains pores which are created due to accumulation of Ar during the growth process. Thus IBSD-SiO$_2$ coatings need to be sufficiently thick to protect the metallic film when forming soft-matter layers, like model membranes, in presence of aqueous solution under physiological relevant buffer conditions.

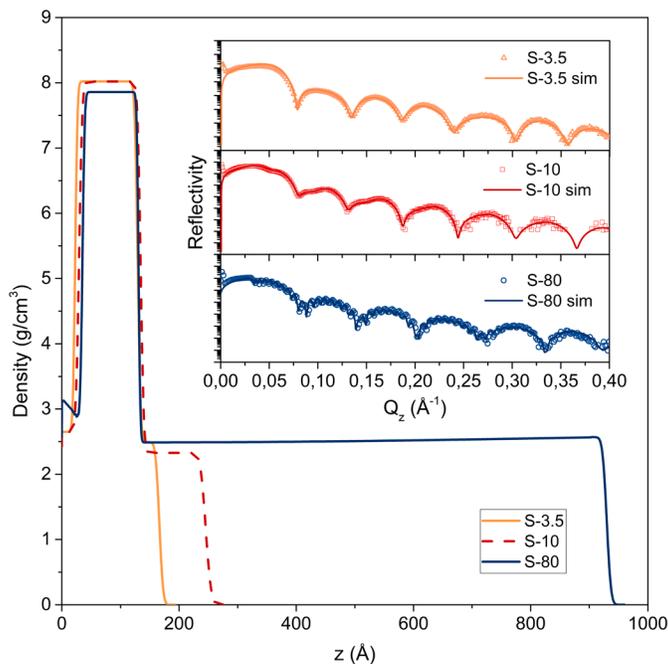

Figure 3. a) XRR curves of S-3.5 (red triangles), S-10 (blue circles), and S-80 (green squares), with corresponding best fit model curves. b) Density profiles of S-3.5 (red solid line), S-10 (blue dashed line), S-80 (green dash-dot line).

All three structures showed pure ferromagnetic behaviour with the coercive force $H_C$=7.5-10.5 Oe and saturation magnetisation $M_S$~1450-1700 emu/cm$^3$ with the lowest value for S-80. Corresponding hysteresis loops are presented in Fig. 4. What is very important, all substrates under the study (including those presented in SI) did not exhibit any significant anisotropy.

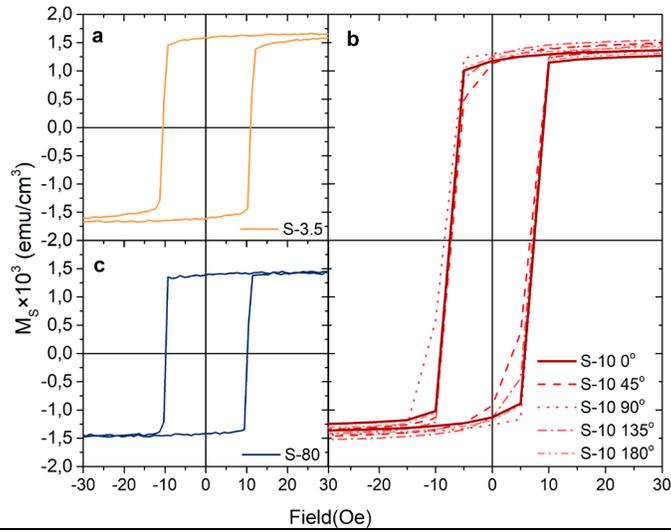

Figure 4. VSM hysteresis loops for a) S-3.5 with $H_C$=10.5 Oe and $M_S$~1650 emu/cm3, b) S-10 with $H_C$=7.5 Oe and $M_S$~1700 emu/cm$^3$ (hysteresis loops were collected at five angular positions in the film plane), and c) S-80 with $H_C$=10 Oe and $M_S$~1450 emu/cm$^3$.

All three substrates were characterised by PNR and showed drastic difference between $R^+$ and $R^-$ reflectivity curves due to the magnetisation of the Fe layer in the external field of 10 mT (Fig. 5). Table 3 shows the structural parameter values obtained from the best fit models for the PNR data, the Fe layer thickness $d_{Fe}$, roughness $\sigma_{Fe}$, as well as the corresponding nuclear and magnetic scattering length densities ($SLD_n$ and $SLD_m$) are very similar for all three substrates. Moreover, the results listed in SI for a larger variety of substrates (Table S2) confirm extremely high level of quality and reproducibility of the Fe layer production. Altogether, our IBSD-based preparation method enables to achieve target thickness $d_{Fe}$ with precision better than 5%, roughness $\sigma_{Fe}$ below 0.5 nm, $SLD_n$ is in agreement with the value for the bulk iron 8.1Å$^{-2}$, $SLD_m$ is 90% of the bulk value 5.5Å$^{-2}$. The later one recalculated into magnetization according to

$$M = C \times SLD_m \left(\frac{emu}{cm^3}\right) \text{ where } C = \frac{2\pi\hbar^2}{\mu_n\mu_0 m_n} = 340 \times 10^6$$

gives magnetization value of 1670 emu/cm$^3$ which is in perfect agreement with the VSM data (see Fig. 4).

The substrates preparation protocol was designed to produce minimum possible roughness of the topmost SiO$_2$ layer. In this respect sample S-80 showed an outstanding value 0.1 nm making system S-80 an ideal platform for the biological NR studies where typical dimension of an object/molecule/structural unit is about 1 nm.

**Table 3.** Structural parameters of MCL samples estimated from PNR data.

| Substrate ID | $\sigma_{sub}$, nm | Fe | | | | SiO$_2$ | | |
|---|---|---|---|---|---|---|---|---|
| | | $d_{Fe}$, nm | $\sigma_{Fe}$, nm | $SLD_n$, Å$^{-2}$ | $SLD_m$, Å$^{-2}$ | $d_{SiO2}$, nm | $\sigma_{SiO2}$, nm | $SLD_n$, Å$^{-2}$ |
| S-3.5 | 0.4 | 10.4 | 0.4 | 8.1 | 4.8 | 3.4 | 0.9 | 3.9 |
| S-10 | 0.6 | 10.4 | 0.4 | 8.1 | 4.9 | 10.0 | 0.9 | 3.9 |
| S-80 | 0.4 | 9.8 | 0.3 | 8.2 | 4.95 | 80.2 | 0.1 | 3.8 |

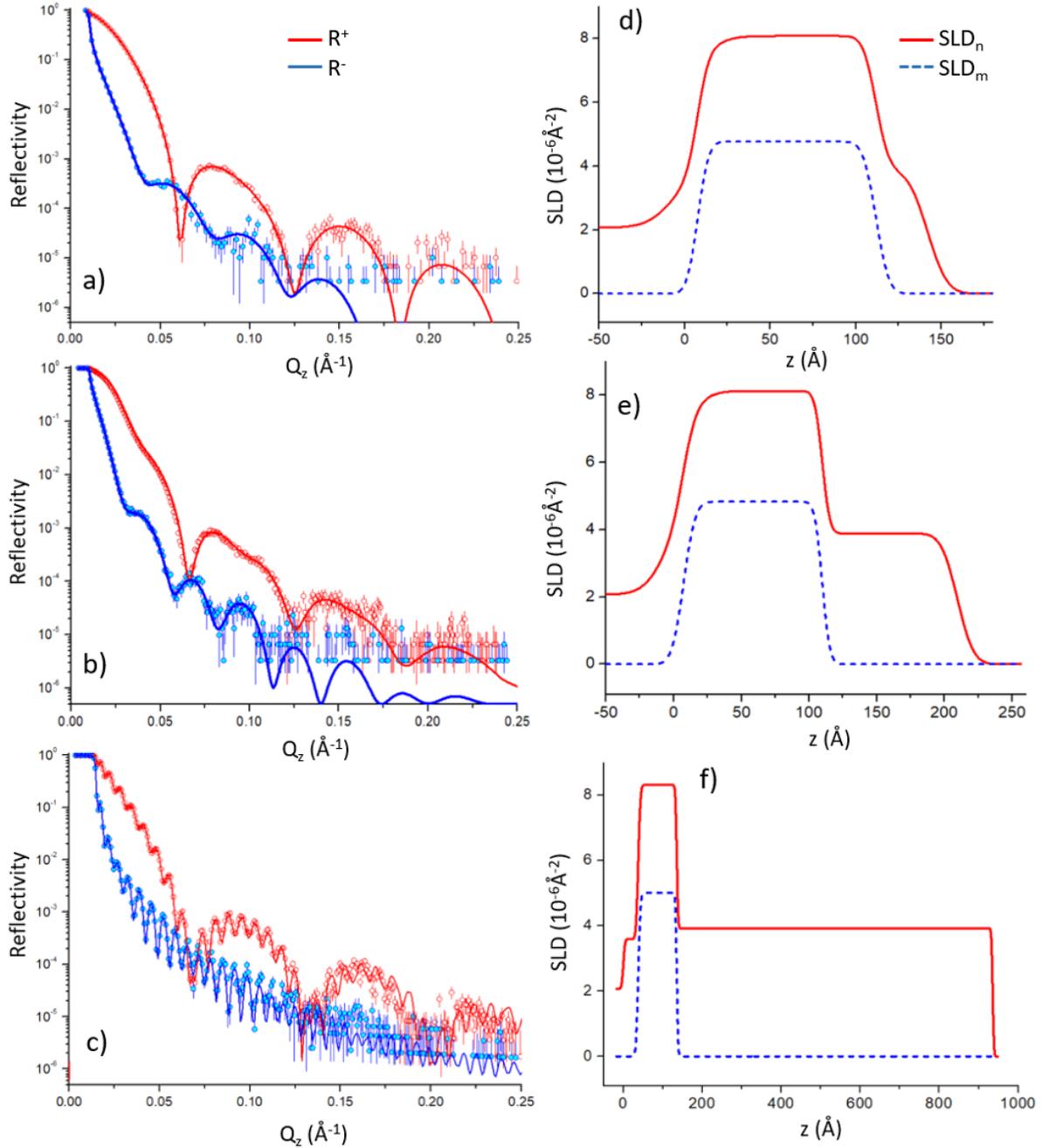

Figure 5. PNR characterisation in external magnetic field of 10 mT, the experimental R+ reflectivity (red circles) and the R- reflectivity (blue circles) with corresponding best fit model curves (red and blue solid lines) for a) S-3.5, b) S-10, and c) S-80. Nuclear $SLD_n$ (red solid line) and magnetic $SLD_m$ (blue dashed line) profiles for d) S-3.5, e) S-10, and f) S-80.

The structure shows stability after three years of shelf storage in ambient conditions, what is proved be measurements of S-10 sample – Figure S6 shows no difference between $R^+$ and $R^-$ reflectivity curves obtained in three-year time difference. PNR data (Figure S7) also provided evidence that magnetisation of the MCL does not change starting from the value of external magnetic field of B=1.5 mT what is again in agreement with the hysteresis data shown in Figure 4. This fact implies that a very low external magnetic field is sufficient to saturate completely the Fe layer what makes construction of the corresponding sample environment extremely easy.

*PNR of the lipid bilayer*

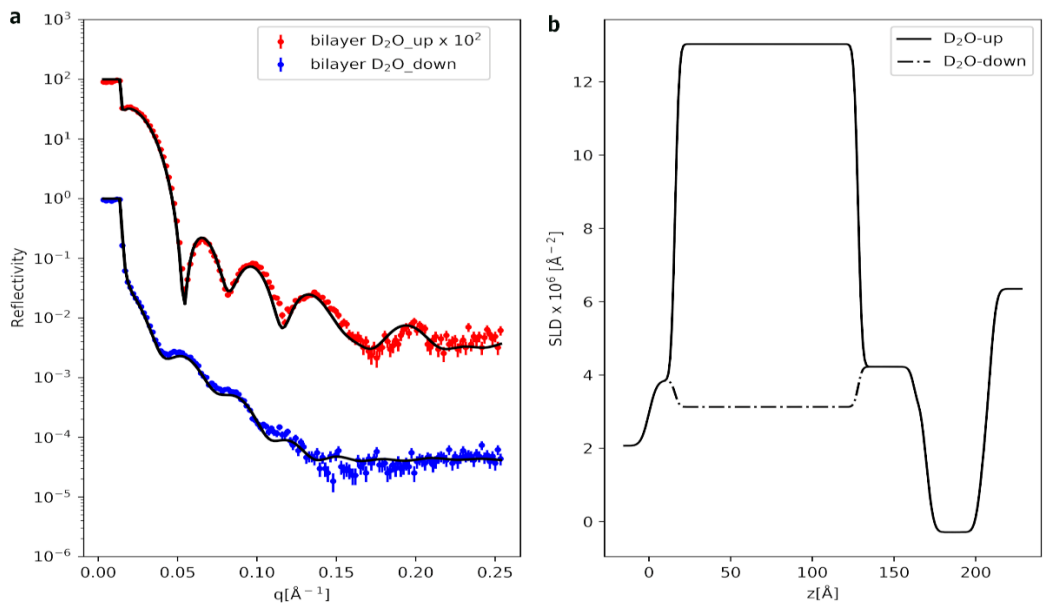

Figure 6. The PNR experimental data and the corresponding fitting curves for the S-3.5 POPC/POPG/Erg (60/30/10) supported lipid bilayer (a). The scattering length density profiles were obtained from the data analysis (b).

The structural parameters for the S-3.5 substrate were in good agreement with those obtained by the other techniques. Figure S8 shows the PNR data for the S-3.5 substrate in contact with $D_2O$ before the lipid deposition. The model used to describe the support was composed of 3 layers: 1) the natural silicon oxide layer (in contact with the Si); 2) the magnetic iron layer; 3) the capping silicon oxide layer (in contact with the $D_2O$). The model parameters were constrained to be the same for the curves referring to spin polarisation up and down expected for the scattering length density of the magnetic layer. The clear difference in the $R^+$ and $R^-$ curves indicates the magnetisation of the thin iron film acting as a magnetic contrast layer.

**Table 3.** Structural parameters were obtained from the analysis of the PNR data collected for the POPC/POPG/Erg supported lipid bilayer on the S-3.5 samples.

| Layer | d, nm | SLD, $10^{-6}$Å$^{-2}$ | σ, nm | Solvent volume fraction |
|---|---|---|---|---|
| $SiO_2$ (natural oxide) | 1.7±0.3 | 3.6 | 0.2±0.2 | - |
| Fe | 11.2±0.4 | $SLD_n$ 8.08 $SLD_m$ 4.08 | 0.3±0.2 | - |
| $SiO_2$ in contact with lipids | 3.7±0.2 | 3.6±0.4 | 0.2±0.1 | 0.15±0.02 |
| Lipid headgroups | 0.7±0.2 | 1.8±0.2 | 0.3±0.1 | 0.20±0.04 |
| Lipid Acyl chains | 3.2±0.03 | -0.23±0.03 | 0.3±0.1 | 0.01±0.01 |

Figure 6 shows the PNR experimental and corresponding fitting curves for the POPC/POPG/Erg supported lipid bilayer on the S-3.5 substrate. The model used to analyse the data included the layers described above for the bare substrate and three additional layers to represent the outer and inner headgroup layers and the intermediate acyl chain region of the lipid bilayer. A preliminary analysis of the experimental data suggested that the bilayer had a symmetric structure. Hence, the two headgroup layers were constrained to have the same structural parameters in the final analysis. This approach reduced considerably the number of fitting parameters in the model. The structural parameters obtained from data analysis are reported in Table 4. Notably, the supported lipid bilayer showed ~100% coverage, as indicated by the very low value for the solvent volume fraction (0.01) in the lipid acyl chain layer. This proves that the Si crystal with the MCL and $SiO_2$ layers has the same performance towards supported lipid bilayer formation as the traditional Si crystals.

**Conclusion**

In conclusion, we introduced a successful method for preparing magnetic contrast layers with $SiO_2$ coatings of low roughness suitable for soft matter PNR studies. To the best of our knowledge, this is the first protocol that enables the production of substrates with sizes compatible with PNR measurements and such low surface roughness. Using the silicon oxide capping layer provides ta suitable surface for the spontaneous deposition of soft matter samples and protects the magnetic layer from degradation in low-corrosive and physiological environments. The produced substrates were characterized by using different complementary techniques, i. e. AFM, XRR, VSM and PNR. As a result, we also found that the magnetic and structural properties of the MCL do not depend on the thickness of top $SiO_2$ layer. As a proof of concept, we used one of the produced substrates to investigate the structure of a relevant model system for the yeast plasma membrane, consisting of a POPC/POPG/Erg bilayer. Interestingly, the sample structure could be efficiently characterised by using the magnetic contrast and avoiding the solvent exchange. Therefore, our method provides a valuable tool for harmless investigation of complex systems in conditions close to their native environment, which has numerous applications in soft matter research.

**Acknowledgements**

The authors express their gratitude to the ILL (Grenoble, France) for providing beamtime at the Super ADAM [24].

**Funding**

The Super ADAM project at Institut Laue-Langevin (ILL, France) is financed by Swedish Research council through grant DNR 2021-00159.

Synuclein to Supported Lipid Bilayers: Positioning and Role of Electrostatics, ACS Chem Neurosci. 4 (2013) 1339–1351. https://doi.org/10.1021/cn400066t.

[6]  A. Luchini, F.G. Tidemand, N.T. Johansen, F. Sebastiani, G. Corucci, G. Fragneto, M. Cárdenas, L. Arleth, Dark peptide discs for the investigation of membrane proteins in supported lipid bilayers: the case of synaptobrevin 2 (VAMP2), Nanoscale Adv. 4 (2022) 4526–4534. https://doi.org/10.1039/D2NA00384H.

[7]  C.F. Majkrzak, K.V. O'Donovan, N.F. Berk, Polarized Neutron Reflectometry, in: Neutron Scattering from Magnetic Materials, Elsevier, 2006: pp. 397–471. https://doi.org/10.1016/B978-044451050-1/50010-0.

[8]  S.A. Holt, A.P. Le Brun, C.F. Majkrzak, D.J. McGillivray, F. Heinrich, M. Lösche, J.H. Lakey, An ion-channel-containing model membrane: structural determination by magnetic contrast neutron reflectometry, Soft Matter. 5 (2009) 2576–2586. https://doi.org/10.1039/b822411k.

[9]  H.-H. Shen, D.L. Leyton, T. Shiota, M.J. Belousoff, N. Noinaj, J. Lu, S.A. Holt, K. Tan, J. Selkrig, C.T. Webb, S.K. Buchanan, L.L. Martin, T. Lithgow, Reconstitution of a nanomachine driving the assembly of proteins into bacterial outer membranes, Nat Commun. 5 (2014) 5078. https://doi.org/10.1038/ncomms6078.

[10] A. Junghans, E.B. Watkins, R.D. Barker, S. Singh, M.J. Waltman, H.L. Smith, L. Pocivavsek, J. Majewski, Analysis of biosurfaces by neutron reflectometry: From simple to complex interfaces, Biointerphases. 10 (2015) 019014. https://doi.org/10.1116/1.4914948.

[11] J.W. Kingsley, P.P. Marchisio, H. Yi, A. Iraqi, C.J. Kinane, S. Langridge, R.L. Thompson, A.J. Cadby, A.J. Pearson, D.G. Lidzey, R.A.L. Jones, A.J. Parnell, Molecular weight dependent vertical composition profiles of PCDTBT:PC71BM blends for organic photovoltaics, Sci Rep. 4 (2015) 5286. https://doi.org/10.1038/srep05286.

[12] A.P. Dabkowska, M. Valldeperas, C. Hirst, C. Montis, G.K. Pálsson, M. Wang, S. Nöjd, L. Gentile, J. Barauskas, N.-J. Steinke, G.E. Schroeder-Turk, S. George, M.W.A. Skoda, T. Nylander, Non-lamellar lipid assembly at interfaces: controlling layer structure by responsive nanogel particles, Interface Focus. 7 (2017) 20160150. https://doi.org/10.1098/rsfs.2016.0150.

[13] M. Khokhlova, M. Dykas, V. Krishnan-Kutty, A. Patra, T. Venkatesan, W. Prellier, Oxide thin films as bioactive coatings, Journal of Physics: Condensed Matter. 31 (2019) 033001. https://doi.org/10.1088/1361-648X/aaefbc.

[14] A. Vorobiev, A. Devishvilli, G. Palsson, H. Rundlöf, N. Johansson, A. Olsson, A. Dennison, M. Wollf, B. Giroud, O. Aguettaz, B. Hjörvarsson, Recent upgrade of the polarized neutron reflectometer Super ADAM, Neutron News. 26 (2015) 25–26. https://doi.org/10.1080/10448632.2015.1057054.

[15] A.Y. Goikhman, S.A. Sheludyakov, E.A. Bogdanov, Ion Beam Deposition for Novel Thin Film Materials and Coatings, Materials Science Forum. 674 (2011) 195–200. https://doi.org/10.4028/www.scientific.net/MSF.674.195.

[16] D. Nečas, P. Klapetek, Gwyddion: an open-source software for SPM data analysis, Open Physics. 10 (2012) 181–188. https://doi.org/10.2478/s11534-011-0096-2.

[17] J.F. Ankner, G.P. Felcher, Polarized-neutron reflectometry, J Magn Magn Mater. 200 (1999) 741–754. https://doi.org/10.1016/S0304-8853(99)00392-3.11